\documentclass[twocolumn,showpacs,amsmath,amssymb]{revtex4}
\usepackage{epsfig}
\usepackage{dcolumn}
\usepackage{bm}
\usepackage{amsmath}
\usepackage{amsfonts}
\usepackage{amssymb}
\usepackage{graphicx}%
\newcommand{\beq}{\begin{eqnarray}}
\newcommand{\eeq}{\end{eqnarray}}

\begin{document}
\title{dc-Response of a Dissipative Driven Mesoscopic Ring.}
\author{Liliana Arrachea$^{1,2}$} 
\affiliation{
$^{(1)}$ Max Planck Institut f\"ur Physik komplexer Systeme, Dresden,
N\"othnitzer Str. 38 D-1187, Germany.\\
$^{(2)}$ Departamento de F\'{\i}sica,
Universidad de Buenos Aires, Ciudad Universitaria Pabell\'on I, (1428) Buenos
Aires, Argentina.}
\pacs{72.10.-d,73.23.-b,73.63.-b}

\begin{abstract}
The behavior of the dc-component of the current along a quantum
loop of tight-binding electrons threaded by a magnetic flux that varies linearly in time  
$\Phi_M(t)=\Phi t$ is investigated. We analize the electron transport in
different kinds of one-dimensional structures bended into a ring geometry:
a clean one-dimensional metal, a chain with a two-band structure and a 
disordered chain.
Inelastic scattering events are introduced through the coupling
to a particle reservoir. We use a theoretical treatment based in 
  Baym-Kadanoff-Keldysh non-equilibrium 
Green functions, which allows us to solve the problem exactly.
 
\end{abstract}
\maketitle

\section{Introduction}
The impressive development of nanoscience places
the detailed understanding of quantum transport in  
mesoscopic systems among the main challenges of 
condensed matter physics.
A rich variety of devices and structures,
including simple metallic wires, as well as
complex molecules,
 where electrons 
are driven by some external force, are the subject of
experimental and theoretical investigation.
The driving field can be established by attached leads at different chemical
potentials, a magnetic flux when the system is bended into a ring
or time-dependent fields. Transport properties, like the conductance
of the system, depend strongly on its microscopic details but they may also
depend on the underlying driving mechanism.   
 
A simple quasi one-dimensional annular system threaded by a magnetic
flux is one of the paradigmatic devices to discuss the fundamentals of 
quantum effects dominating the electron transport. 
Several outstanding experiments \cite{ab1,ab11,ab12} and 
theoretical works \cite{ab2,imb}
have been devoted 
to the investigation of phenomena related to the Aharanov-Bohm effect
in rings enclosing a static magnetic field. There is, however, relatively
less literature related to the case where the field changes in time.  
Among the latter category of problems, a very interesting example
corresponds to that of a magnetic flux with a linear dependence on time:
$\Phi_M(t)=\Phi t$, which induces a constant electric field along the ring. 

This problem was discussed by  B\"uttiker, Imry and
Landauer in the early times of the theory of quantum transport \cite{ibl}.
Their motivation was to search alternative schemes to calculate the 
conductance of a system to that proposed by Landauer \cite{lan},
where the sample is placed between two reservoirs at different 
chemical potentials. As first discussed in that seminal work, such device 
isolated from the external world, is not appropriate to observe a dc-current
since the electrons inside move coherently displaying Bloch oscillations and
producing a pure ac-response. In a subsequent work 
Landauer and B\"uttiker \cite{bl} showed that when a dissipative 
mechanism is
added to that system a dc-current is established.  B\"uttiker also
showed that a concrete element to introduce inelastic scattering events
and dissipation is a lead connecting the loop to a particle 
reservoir \cite{but}. 
Soon later, Lenstra and van Haeringen suggested that a dc-current
can be generated in that device by elastic scattering introduced
by weak disorder \cite{lh}. The possibility of ``resistive'' behavior 
originated in pure elastic scattering processes generated 
a series of interesting discussions and criticisms 
\cite{lan1,lan2,gef1,gef2,brou,ao,av}.  
Most of the ideas of these works are
based on adiabatic descriptions where
 the energy levels of the ring define minibands in a parametric
representation as periodic functions of the flux.
Within that adiabatic framework, scattering processes form small gaps between 
the so defined minibands while the time dependence of the field 
gives rise to Zener tunneling across them. 
In Refs. \cite{gef1,gef2} interesting arguments emphasizing the concept
of localization in the energy space
as a consequence of disorder have been proposed against the possibility
of dc-response without dissipation.

The discussion of the role of dissipative effects in the
driven ring was introduced on the basis of
a phenomenological equation
of motion that describes the relaxation via inelastic scattering
processes (ISP) of 
the time-dependent occupation of a miniband \cite{bl}.
The main argument was that the dc-response should vanish in the limit
of vanishing and strong relaxation, while it should peak
at some intermediate regime. 
For small electric fields, the behavior of the current is also found to 
follow a linear,
ie. Ohmic-like, dependence as a function of the induced electromotive force 
(emf). 
These notable predictions have not been examined in more detail during many 
years.
Quite recently, the effect of inelastic scattering in the dc-behavior 
 of that system has been studied \cite{liu}. In that work a pure ``clean'' 
system
is considered, (ie without
any kind of elastic scattering). Non-Ohmic behavior has been found in the
dc-current vs emf characteristic curve within the limit of small ISP,
 while the dc-current is found
to decrease continuously as the strength of the ISP increases.
The fact that no tendency towards a vanishing component of the dc-current is 
observed
as the dissipation tends to be suppressed is rather surprising,
since the limit of pure Bloch oscillations in the isolated system 
seems not to be recovered.

In this work we consider a ring  threaded by a flux with a linear time 
dependence and we analyze in detail the role of dissipation in that 
system. We study the clean one-band system as well as the effect of two 
different kinds of elastic
scattering processes: a periodic 
 potential with a two-sublattice profile
and a random potential.
 The effect of inelastic scattering is modeled by means of the coupling to
a particle reservoir through an external lead.
We use a recently proposed 
 theoretical treatment based on Baym-Kadanoff-Keldysh
Green functions \cite{lili} which enables a full out-of-equilibrium
description of the time-dependent problem, without introducing any kind of 
adiabatic
assumptions or approximations. As far as no many-body interactions are
 considered, that treatment leads to the exact solution of this problem. 
Thus, one of the aims of this work is to analyze the case of the disordered
driven ring in the limit of weak ISP, which has focused 
several efforts but has never been tackled with an exact method.
 The formalism based on Green
functions is particularly appealing to deal with the coupling to external leads
and reservoirs. These elements can be introduced by means of physical models 
and
 they can be concisely represented by self-energies, avoiding further
assumptions 
about the
boundary conditions. We consider two different electronic models for 
the lead: 
one is described by a constant density of states and a very wide band,
while the other is represented by a semi-infinite
tight-binding chain.
 We
show that the dc-response does not depend qualitatively on the latter details.
In all the cases, our results
indicate  that the dc-current tends to a vanishing 
value as the
coupling to the reservoir goes to zero, displays a maximum and then decreases
as that coupling becomes stronger. 
For clean systems, the position of the maximum
shifts towards very low ISP strengths as the emf decreases and as the 
length of the 
ring increases. 

 The system
considered here, actually lies in the category of the so called {\em ratchet}
problems, since the induced electric field by itself is not able to 
produce a dc-response, needing the aid of some additional rectification
mechanism.
This motivates the analysis of the transport properties of this system
in the framework of recent symmetry arguments proposed to examine
rectification processes in  ratchet systems
in the presence of 
 time-periodic fields \cite{ser1,ser2,ser3}.
Such arguments provide further support to the idea that dissipation is
an essential ingredient for this system to have a dc-current irrespectively
the particular nature of the elastic scattering along its circumference.

The paper is organized as follows. In section II we present the model 
and the theoretical approach to analyze the behavior
of the current.
In section III we present the results. Section IV is devoted to summary and 
conclusions.

\section{Theoretical treatment}
\subsection{The model}
\begin{figure}
\includegraphics[width=50mm,angle=-90]{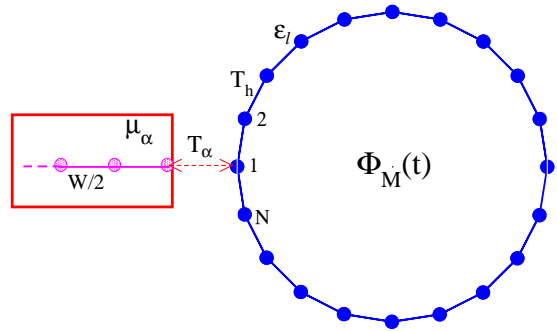}
  \caption{(Color on line) Scheme of the setup. The system indicated in the red box defines the
reservoir with chemical potential $\mu_{\alpha}$.}
\label{fig0}
\end{figure}
We consider the device sketched in Fig. \ref{fig0}. It consists
in a ring threaded by a magnetic flux with a dependence on time
of the form $\Phi_M(t)=\Phi t$, in contact to 
a particle reservoir 
with a chemical potential $\mu_{\alpha}$ through an external lead. 
The full system 
is described by the following Hamiltonian:
\begin{equation}
H= H_{ring}(t)+H_{\alpha}+H_{1 \alpha},
\label{e1}
\end{equation}
where the first term, representing the ring, depends
explicitly on time due to the presence of the 
time-dependent flux.
We consider noninteracting
spinless electrons  described  by a tight binding model with $N$ sites
and lattice constant $a=L/N$. We also consider the possibility of
elastic scattering in the system, described by a 
local energy with a profile $\epsilon_l$ with $l=1,\ldots, N$. The Hamiltonian
is
\begin{equation}
H_{ring}=-T_h\sum_{l=1}^{N} (  e^{-i \phi t} c^{\dagger}_l c_{l+1} +  
 e^{i \phi t} c^{\dagger}_{l+1} c_l) + 
\sum_{l=1}^{N} \epsilon_l c^\dagger_l c_l ,
\label{e2}
\end{equation}
with the periodic condition $N+1 \equiv 1$.
The time-dependent phase $\phi t$  attached
to each link,  with $\phi=\Phi/(\Phi_0 N)$,
being $\Phi_0= hc/e$, 
accounts for the  presence of the external magnetic flux.

The term $H_{\alpha}$ describes  the reservoir. 
We consider two models of noninteracting
electrons with a bandwidth $W$ and a
 chemical potential $\mu_{\alpha}$ for this system: 
(i) a wide-band model,
defined by a constant density of states and a very large
$W$, and 
(ii) a semi-infinite tight binding chain with hopping amplitude 
$W/2$, which corresponds to a semicircular  
density of states
$\rho_{\alpha}(\omega)=4 \Theta(|\omega|-W) \sqrt{W^2-\omega^2}/W^2$.

The last term of $H$,
\begin{equation}
H_{1 \alpha }= -T_{\alpha} 
(c^{\dagger}_1 c_{\alpha} +c^{\dagger}_{\alpha} c_1 ), 
\label{e3}
\end{equation}
represents the connection between the ring and the reservoir.

\subsection{Dynamical equations and symmetry properties}
In Refs. \cite{ser1,ser2,ser3} an interesting
 connection has been suggested between the
rectification properties of the system and the underlying symmetries 
of the equations of motion in several ratchet problems. 
The  
main idea is that the symmetries of the equation of
motion that change the sign of the velocity would lead to a vanishing
dc-component of the current.  
We now turn to follow the lines suggested in those works to
carry out a similar symmetry analysis in our problem.

Let us first consider the simpler case  
of the ring isolated from the reservoir, the problem is 
described by  $H_{ring}$ alone. 
 For sake of simplicity in
the notation we shall adopt a system of units
where $\Phi_0=1,\;\hbar=1$.
The relevant equation of motion
is the Schr\"odinger equation
\begin{equation}
-i \frac{\partial}{\partial t}\psi_l(t)= \varepsilon_{lk}(t)\psi_k(t),
\label{sch}
\end{equation}
being $\varepsilon_{lk}(t)= -T_h (e^{-i \phi t} \delta_{k,l+1}+ 
e^{i \phi t} \delta_{k,l-1})$. 
We have adopted the Einstein summation rule over site indexes.
The current in the isolated ring results
\begin{equation}
J^{isol}_{l,l+1}(t)=e T_h \mbox{Im}[e^{-i \phi t} \psi^*_l(t) \psi_{l+1}(t) ].
\label{isolc}
\end{equation}
The matrix elements of the Hamiltonian satisfy
\begin{equation}
 \varepsilon_{lk}^*(-t)=\varepsilon_{lk}(t)=\varepsilon^*_{kl}(t). 
\label{prop}
\end{equation}
Hence, the inversion 
$t \rightarrow -t$ followed by the complex conjugation of (\ref{sch})
leads to $\psi^*_l(-t)=\psi_l(t)$  which, when replaced
in (\ref{isolc}), results in the following property of the current
\begin{equation}
J^{isol}_{l,l+1}(-t)=-J^{isol}_{l,l+1}(t).
\label{isol}
\end{equation}

Since in our reasoning we have not assumed any particular form for the   
energy profile $\epsilon_l$, the criterion of Refs.
\cite{ser1,ser2,ser3} rules
out the possibility of a dc-current as far as time-reversal symmetry
is preserved in the device irrespectively the particular model
assumed for the elastic
scattering processes. Another symmetry operation that changes the sign of 
the current is spatial reflection, however, the latter is not a
symmetry of the Schr\"odinger equation (\ref{sch}). 

In the more general case,
when inelastic scattering processes are considered, 
the relevant equation of motion is the Dyson equation for the 
Green function. In the framework of Baym-Kadanoff-Keldysh formalism the latter 
has a matrix
form and one must work with two independent
Green functions: the retarded Green function,
\begin{equation}
G^R_{i,j}(t,t')=-i \Theta (t-t')
 \langle [ c_i(t), c^{\dagger}_j(t') ] \rangle,
\end{equation}
and the lesser Green function
\begin{equation}
G^<_{i,j}(t,t')=i \langle  c^{\dagger}_j(t') c_i(t) \rangle.
\end{equation}  
The latter determines the mean values of the observables. 
In particular, the current through a bond $\langle l, l+1 \rangle$ is 
written as
\begin{equation}
J_{l,l+1}(t)=2e Re[ T_h e^{-i\phi t} G^{<}_{l+1, l}(t,t)].
\label{e4}
\end{equation}

The Dyson equation for the matricial Green function leads to
the equations of motion 
for the retarded and lesser components \cite{ram}
\begin{eqnarray}
& &-i \frac{\partial}{\partial t'} G^R_{ij}(t,t') - G^R_{ik}(t,t')
\varepsilon_{kj}(t') = \delta(t-t') \delta_{ij} \nonumber\\
& &+ \int dt'' \, G^R_{ik}(t,t'') \Sigma^R_{kj}(t'',t'),
\nonumber\\
& &-i \frac{\partial}{\partial t'} G^<_{ij}(t,t')  - G^<_{ik}(t,t')
\varepsilon_{kj}(t') = \nonumber\\
& & 
\int dt'' [ G^R_{ik}(t,t'')
\Sigma^<_{kj}(t'',t')  \nonumber\\
& &+ 
 G^<_{ik}(t,t'') \Sigma^A_{kj}(t'',t') ]
\;,
\label{dyson-K2}
\end{eqnarray}
while the  evolution for the lesser component at
equal times is given by
\begin{eqnarray}
& & i \, \frac{d G^<_{ij}(t,t)}{dt} -  G^<_{ik}(t,t)
\varepsilon_{kj}(t) + \varepsilon_{ik}(t) G^<_{kj}(t,t)   \nonumber \\
& &=
\int dt'' \;
[ \; \Sigma^R_{ik}(t,t'') G^<_{kj}(t'',t) +
\Sigma^<_{ik}(t,t'') G^A_{kj}(t'',t) 
\nonumber \\
& & 
-G^R_{ik}(t,t'')
\Sigma^<_{kj}(t'',t)  -
 G^<_{ik}(t,t'') \Sigma^A_{kj}(t'',t)\;
]
\; ,
\label{border}
\end{eqnarray}
being $G^A_{i,j}(t,t')=[ G^R_{j,i}(t',t)]^*$ and
$\Sigma^A_{i,j}(t,t')=[ \Sigma^R_{j,i}(t',t)]^*$.

In the Hamiltonian limit, $\Sigma^R_{i,j}(t,t')=\Sigma^<_{i,j}(t,t')
\equiv 0$, in the absence of many-body interactions. 
Making use of the property of the matrix elements of the
Hamiltonian (\ref{prop})
and of the symmetry property of the Green function
 $G^<_{j,i}(t,t')=-[G^<_{i,j}(t',t)]^*$,
the property (\ref{isol}) for the time-inversion operation
 on the current is recovered.

It is easy to verify that any non-vanishing self-energy
correction that fails
to satisfy the properties
$\Sigma^<_{ik}(t,t')= \Sigma^>_{ik}(t,t')\equiv 0$,
$\Sigma^R_{ik}(t,t')= s_{ik}(t) \delta(t-t')$,
with $s_{ij}(t)= s^*_{ji}(t)=  s_{ji}(-t)$,
breaks time-reversal symmetry in the equations of motion.
In particular, any self-energy represented by kernels of the form
\begin{eqnarray}
\Sigma^R_{ik}(t,t')&=& -i \Theta (t-t')\int\frac{d\omega}{2\pi} 
\{ -2 \mbox{Im}[\Sigma^R_{ik}(\omega)] \}e^{-i \omega (t-t')},\nonumber\\
\Sigma^<_{ik}(t,t')&=&  i \int\frac{d\omega}{2\pi} f(\omega)
\{ -2 \mbox{Im}[\Sigma^R_{ik}(\omega)] \}e^{-i \omega (t-t')},
\label{sigmas}
\end{eqnarray}
with $\mbox{Im}[\Sigma^R_{ik}(\omega)] \neq 0$, 
being $f(\omega)$ the Fermi function, will break time-reversal symmetry
in the equations of motion for the Green functions and will break the
property (\ref{isol}) for the time-dependent current.

In our problem,
the effect of the exchange of particles and energy between the
mesoscopic system and the reservoir
can be exactly 
written in terms of a self-energy correction at the site $l=1$ 
of the tight-binding ring \cite{lili,jau,past}.
 The retarded and lesser components of this self-energy are  
\begin{eqnarray}
\Sigma^R_{ik}(t-t')
&=&\delta_{i,1}\delta_{k,1}\Sigma_{1 }^R(t-t^{\prime})\nonumber \\
&=& -i \Theta(t-t')|T_{\alpha}|^2 \nonumber \\
& & \times \int \frac{d\omega}{2\pi} \rho_{\alpha}(\omega) e^{-i\omega(t-t')}
\delta_{i,1}\delta_{k,1},\nonumber\\
\Sigma^<_{ik}(t-t')&=&\delta_{i,1}\delta_{k,1}\Sigma_{1 }^<(t-t^{\prime})\\
&=& i |T_{\alpha}|^2
\int \frac{d\omega}{2\pi} f(\omega)\rho_{\alpha}(\omega) e^{-i\omega(t-t')}
\delta_{i,1}\delta_{k,1}, \nonumber
\label{e5}
\end{eqnarray}
where 
$\rho_{\alpha}(\omega)$ is the density of states of the reservoir
and the Fermi function is $f(\omega)=1/(e^{\beta(\omega-\mu_{\alpha})}+1)$.
In our calculations we consider zero temperature, i.e.
 $f(\omega)=\Theta(\omega-\mu_{\alpha})$. 
The wide-band model
leads to a constant imaginary retarded self-energy,
$\Sigma_{1}^R(\omega)=i \sigma$, and
$\Sigma_{1}^<(\omega)= i f_{\alpha}(\omega) \sigma$,
being $\sigma=|T_{\alpha}|^2 \pi/W$.
The dissipative nature of the coupling to the reservoir manifests
itself in the fact that
 $\Sigma_{1}^R(\omega)$ has a finite imaginary part. As discussed 
above, this breaks time-inversion symmetry in the dynamical equation,
removing the inversion of the current under this symmetry operation.
This satisfies the criterion  of Refs. \cite{ser1,ser2,ser3}, 
namely, a non-vanishing net current is possible when symmetries of the equation of motion 
leading to an inversion of the time-dependent current are broken.

\subsection{Evaluation of the Green functions and the dc-current.}
The formalism leading to the evaluation of the current along
the ring has been presented in Ref. \cite{lili}.
In this subsection we summarize the main equations 
 and we defer the reader to this work and references
therein for further details.

Following Ref.\cite{lili}, 
it is convenient to perform a gauge transformation in 
the fermionic operators of the Hamiltonian (\ref{e2}),
\begin{equation}
 c_n=\exp[i n \phi t] \overline{c}_n,
\label{gauge}
\end{equation}
according to which 
the Green function for the positions $m,n$ on the ring
transforms as
\begin{equation} 
G^R_{m,n}(t, t^{\prime})=\exp[i \phi (m t - n t^{\prime})]  
\overline{G}^R_{m,n}(t, t^{\prime}).
\end{equation} 
Defining the Fourier transform
\begin{equation}
G^R_{m,n}(t,t')=\int \frac{d\omega}{2 \pi} G^R_{m,n}(t,\omega) 
e^{-i\omega(t-t')},
\end{equation}
the resulting equation for the retarded Green function
is
\begin{eqnarray}
\overline{G}^R_{m,n}(t,\omega)&=&
G^0_{m,n}(\omega)-
\overline{G}^R_{m,N}(t,\omega+\Phi)T_{N1}(t)G^0_{1,n}(\omega)
\nonumber\\
& &
-\overline{G}^R_{m,1}(t,\omega-\Phi)T_{1N}(t)G^0_{N,n}(\omega),
\label{a1}
\end{eqnarray}
with
 \begin{eqnarray} 
 & &\overline{G}^R_{m,1}(t,\omega)+
\overline{G}^R_{m,N}(t,\omega+\Phi) 
T_{N1}(t) G^0_{1,1}(\omega)\nonumber\\
& &+ \overline{G}^R_{m,1}(t,\omega-\Phi) 
T_{1N}(t) G^0_{N,1}(\omega)=G^0_{m,1}(\omega)\nonumber \\ 
& &\overline{G}^R_{m,N}(t,\omega)+\overline{G}^R_{m,N}(t,\omega+\Phi) 
T_{N1}(t) G^0_{1,N}(\omega)\nonumber\\  
& &+ \overline{G}^R_{m,1}(t,\omega-\Phi) 
T_{1N}(t) G^0_{N,N}(\omega)=G^0_{m,N}(\omega), 
\label{a2} 
\end{eqnarray}
where $T_{1N}(t)=[T_{N1}(t)]^*=T_h e^{i \Phi t}$.   
For each time $t$, the solution of the above set of linear equations
provides the complete exact solution of the problem. 

The equilibrium Green function $G^0_{m,n}(\omega)$ corresponds to the
problem of an open chain in contact to the reservoir. 
It is obtained from the solution of the Dyson equation
\begin{equation}
G^0_{m,n}(\omega)= g^0_{m,n}(\omega) + G^0_{m,1}(\omega) \Sigma^R_1(\omega)
g^0_{1,n}(\omega),
\label{a3}
\end{equation}
where 
\begin{equation}
g^0_{m,n}(\omega)=
\sum_{\nu=1}^N A^{\nu}_{m} A^{\nu}_{n} \frac{1}{\omega -E_\nu +i \eta}
\label{a4}
\end{equation}
is the Green function of the chain isolated from the
reservoir. It can be expressed 
in terms of the eigenvalues $E_{\nu}$ and eigenvectors
$|\nu \rangle= \sum_l A^{\nu}_l |l\rangle$ of the Hamiltonian
\begin{equation}
\overline{H}_0=
-T_h\sum_{l=1}^{N-1} 
( \overline{c}^{\dagger}_l \overline{c}_{l+1} +  
\overline{c}^{\dagger}_{l+1} \overline{c}_l) 
+\sum_{l=1}^{N} (V_l+\epsilon_l) \overline{c}^{\dagger}_l \overline{c}_l,
\label{e8}
\end{equation} 
where $V_l= \phi l$ is the scalar potential
due to the induced electric field.

The set (\ref{a1}) and (\ref{a2}) describes the process of the closing of the 
biased chain by means of an effective time-dependent hopping $T_{1N}(t)$
that {\em pumps} electrons  with a frequency $\Phi$ through the
bond $\langle 1N \rangle$.
This set  involves, in principle, an infinite number of
equations. In the numerical procedure, upper and lower energy 
cutoffs $\Lambda_1, \Lambda_2$ are chosen  such that 
$\Lambda_1<E_{\nu}<\Lambda_2, \;\forall \nu$.

Starting from its definition
(\ref{e4}), the current along the ring can be written
\begin{eqnarray}
J_{l,l+1}(t)&=& 2e T_h \int \frac{d \omega}{2 \pi}  
\mbox{Re}[\Sigma_1^<(\omega) \nonumber \\
& &\times \overline{G}^R_{l+1,1}(t,\omega+\phi) [\overline{G}^R_{l,1}(t,\omega+\phi)]^*],
\label{curt}
\end{eqnarray}
where we have used
$G^A_{1,l}(t,t')=[G^R_{l,1}(t',t)]^*$.
Therefore, the evaluation of the time-dependent current 
is reduced to the evaluation of 
$\overline{G}^R_{l,1}(t,\omega)$ obtained from (\ref{a2}).
This quantity displays an oscillatory behavior as a function of $t$,
with the period $\tau_B=2\pi/\Phi$ of the Bloch oscillations.
The dc-component does not depend on the bond $\langle l, l+1\rangle$
chosen for the calculation and it is defined as
\begin{equation}
J_{dc}=\frac{1}{\tau_B}\int_0^{\tau_B} dt J_{l,l+1}(t).
\label{curdc}
\end{equation}

\section{Results}
This section is devoted to analyze the behavior of the dc-current (\ref{curdc})
as a function of the 
induced emf $\Phi$, the strength of ISP and the chemical potential 
$\mu_{\alpha}$. 
The strength of ISP is related to the degree of coupling between the ring
and the reservoir. In the case of the wide band model, the parameter $\sigma$
sets that measure. In the model with
a semicircular density of states, that measure is given by 
$T_{\alpha}^2/W$. 
In our calculations, we fixed $W=4T_h$ and changed 
$T_{\alpha}$. All energies will be expressed in units of the 
hopping parameter $T_h$. We shall analize three different 
energy profiles $\epsilon_l$ for the tight-binding model
that define a clean one-band chain, a chain with a 
two-sublattice structure and a random potential. 

\subsection{The clean one-band ring}
This case corresponds to $\epsilon_l=0, l=1,\ldots,N$ in (\ref{e2}).
In the limit of vanishing dissipation (the ring isolated from the 
reservoir) the system is described by the Hamiltonian $H_{ring}(t)$ alone,
the problem has time-reversal symmetry  and a vanishing
dc-current is expected. In fact, for $\epsilon_l=0$, $H_{ring}(t)$
 can be easily solved by
performing a Fourier transform to the $k$-space. The retarded
Green function can be calculated analytically, resulting
$G^R_{m,n}(t,t')=-i\Theta(t-t')
\sum_k \exp\{ik (m-n)\}
\exp\{-i\int_t^{t'} ds \epsilon_k(s)\}$,
with $\epsilon_k(s)=-2T_h \cos(k+\phi s)$, being 
$k=2 n \pi/L, n=0, \ldots, N$.
The current can also be obtained
analytically, resulting 
\begin{equation}
J(t)=T_h \sum_{k} \sin(k+\phi t).
\end{equation}
where the summation extends over the set of occupied $k$-states.

\begin{figure}
\includegraphics[width=65mm,angle=-90]{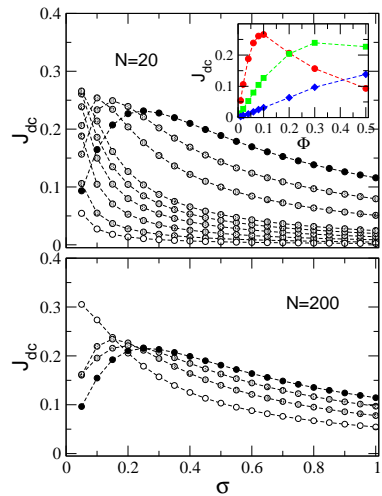}
  \caption{(Color online) dc-current as a function of $\sigma$ in a ring with $N=20$ sites
for different values of the induced electric field 
$\Phi=0.01,0.02,0.04,0.06,0.08,0.1,0.2,0.3,0.5$ (upper panel) and with $N=200$ for
$\Phi=0.2,0.3,0.4,0.5$ (lower panel). 
The chemical potential of the reservoir is $\mu_{\alpha}=-1$. The plots corresponding to
the lowest and the highest $\Phi$ are drawn in white and black symbols, respectively, while
grey symbols correspond to intermediate values.
Inset: 
$J_{dc}$ vs $\Phi$ characteristic curve 
for different strengths of ISP
corresponding to $\sigma=0.05,0.2,0.8$ (red circles, green squares and blue diamonds).}
\label{fig1}
\end{figure}

The behavior of the dc-current in the coupled system
at a fixed chemical potential $\mu_{\alpha}$
for different strengths of ISP and $\Phi$ is analyzed in
Fig. \ref{fig1}. A wide band model was assumed for the reservoir in these calculations. 
The dc-current displays a maximum which shifts to lower values of
$\sigma$ as $\Phi$ decreases. At low $\sigma$,
the imaginary part of $G^0_{m,n}(\omega)$ 
evolve towards a sequence of delta functions and the 
Green functions $\overline{G}^R_{m,n}(t,\omega)$ also develop 
a structure of peaks that get sharper.   
From the practical point of view, this turns harder 
the evaluation of the integral in $\omega$ (\ref{curt}), preventing
us from exploring the range of $\sigma/T_h<0.05$. However, we think that
the range considered is enough to infer the trend towards 
$\sigma \rightarrow 0$. 
The characteristic curves 
$J_{dc}$ vs $\Phi$ are shown in the inset for particular values of
$\sigma$ within  the regimes with low, intermediate and strong coupling to
the reservoir. 
In the latter limit, the dc-current remains linear within a wide range 
of $\Phi$ while for weak coupling, the departure from linear
response takes place at small $\Phi$. 

To study the origin of this peculiar behavior we must 
analyze the structure of the equations (\ref{a1})
and (\ref{a2}). 
At each time $t$, the Green function $\overline{G}^R_{m,n}(t,\omega)$
contains a combination of a large number of $\omega$-components 
(separated in $\Delta \omega = \Phi$)
of the Green function (\ref{a3}). In the weak coupling
limit, $G^0_{m,n}(\omega)$ is sizable only within a neighborhood of
the $N$ frequencies $E_{\nu}$. Therefore, a {\em resonant} combination of 
components at different frequencies is achieved when $\Delta \omega \sim
\Delta E_{\nu} $, the latter being the mean energy separation 
between two eigenenergies, 
$\Delta E_{\nu}=\langle E_{\nu+1}-E_{\nu} \rangle \sim 4T_h/N$.
Whether such interference 
will be constructive or destructive and give rise to a large or a small
component of the dc-current is a question without obvious answer, 
particularly in the
case of small $\Phi$, where the number of coupled frequencies is
very large. On general grounds, one expects that such an effect would
strongly depend on the underlying symmetries of the model. 
Related discussions have been recently presented in problems of
pumping introduced by microwave fields \cite{leh,han,mos} where the
Floquet representation of the wave function is used, leading to
a structure of the solution containing a mixing of frequencies that differ
in the frequency of the pumping field.
In our problem, it becomes apparent that such interference is constructive
regarding the magnitude of $J_{dc}$
and for this reason, in the weak coupling limit, the current grows linearly as
a function of $\Phi$, reaching a maximum at $\Phi \sim \Delta E_{\nu}$.
The decrease for $\Phi >> \Delta E_{\nu}$ can be understood by noting
that the number of resonances is approximately $4T_h/\Phi$,
becoming smaller as $\Phi$ increases.
The maximum of $J_{dc}$ as function of $\sigma$ can be also
 interpreted in terms of combination of components of the Green
function from different frequencies. In fact, the effect of increasing
ISP is to spread the spectral weight
of the peaks of $G_{m,n}^0(\omega)$. Thus, for strong coupling the
interference involves a large number of frequencies
but with a low (approximately constant)
weight. For lower ISP, the amplitudes can be large
for frequencies close to $E_{\nu}$ but tend to be vanishingly
small in between. 
The result is that there is a maximum in $J_{dc}$ at some
strength of ISP that seems to scale as $\propto \Phi$,
at least,
within the regime $\Phi>\Delta E_{\nu}$.
On the side of strong coupling, where
the resonant effects are highly smoothed,
the resulting dc-current is relatively smaller (consistent with 
the idea that resistance increases with ISP)
and behaves linearly within a larger range of $\Phi$. 

The effect of the length of the chain is also analyzed
in Fig.\ref{fig1}. The first issue to note is that in the $N=200$ sites
ring (see lower panel of Fig. 1, the dc-current for $\Phi=0.2$
remains growing 
for the smallest $\sigma$ considered.
Instead, for $N=20$ and the same parameters, the current 
decreases down from its maximum (cf upper panel of Fig. \ref{fig1}). 
This behavior is consistent
with the picture of constructive resonances: The condition 
$\Phi > \Delta E_{\nu}$
is achieved in this case at lower $\Phi$, while the larger number of peaks
increases the probability of resonances and the current remains large, 
within a larger range of $\Phi$. For larger
fields and for larger $\sigma$, the qualitative and
quantitative behavior is essentially the same as that observed in the smaller
ring.

\begin{figure}
\includegraphics[width=65mm,angle=-90]{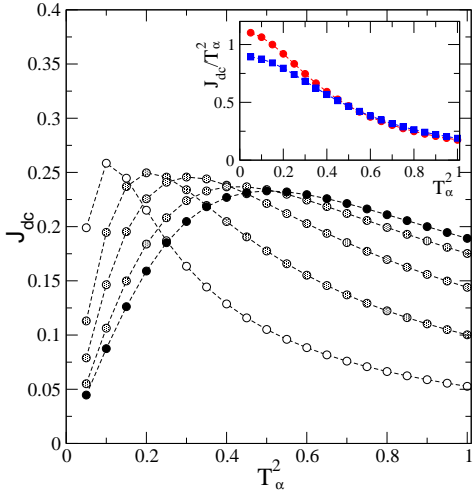}
  \caption{(Color online) dc-current as a function of the strength of ISP for a model with
a semicircular density of states with
bandwidth $W=4$.
Different plots correspond to 
$\Phi=0.1,0.2,0.3,0.4,0.5$. The plots in white and black symbols correspond to the lowest and the 
highest $\Phi$, while the grey ones correspond to the intermediate values. 
The chemical potential of the reservoir is $\mu_{\alpha}=-1$.
Inset: Detail of the behavior of $J_{dc}$ for low ISP. 
Red circles and blue squares correspond to $\Phi=0.4, 0.5$, respectively.
The behavior is consistent with
$J_{dc} \propto  T^2_{\alpha}$ when $T^2_{\alpha} \rightarrow 0.$
}
\label{fig2}
\end{figure}
Regarding the effect of the model for the reservoir, 
it becomes clear from the results of 
 Fig. \ref{fig2} that it does not play any relevant qualitative
role since we can identify in these plots the same features observed in Fig. \ref{fig1}.
In Ref. \cite{leh} an interesting connection have been found between the behavior of
$J_{dc}$ as a function of $T^2_{\alpha}$ when $T^2_{\alpha} \rightarrow 0$ and the underlying
symmetries of the Hamiltonian.
The plots of Fig. \ref{fig2} corresponding to the two highest emfs ($\Phi=0.4,0.5$), where the
decrease to a vanishing $T^2_{\alpha}$ can be cleanly
captured wihin the shown range of $T^2_{\alpha}$, suggest a $J_{dc} \propto T^2_{\alpha}$
behavior as  $T^2_{\alpha} \rightarrow 0$. This is even more clear in the plot
of $J_{dc}/ T^2_{\alpha}$ vs $T^2_{\alpha}$ shown in the inset. This dependence can
be understood by noting that when $ \Delta E_{\nu} + \Phi >> T^2_{\alpha}/W$, the
retarded Green functions entering the evaluation of $J_{l,l+1}(t)$ (see Eq. \ref{curt}),
can be replaced by the ones for the uncoupled ring (corresponding to $T^2_{\alpha}=0$).
Hence, for small enough $T^2_{\alpha}$, the time-dependent current and also the dc-component $J_{dc}$
become linear in $T^2_{\alpha}$. This is in perfect agreement with the analysis done in 
Ref. \cite{leh} for a linear device pumped with a laser in configurations with broken time-reversal
symmetry.

\begin{figure}
\includegraphics[width=65mm,angle=-90]{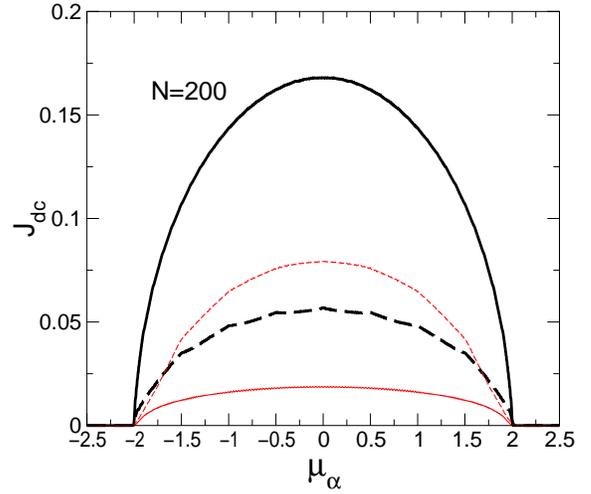}
  \caption{(Color online) dc-current as a function of the chemical potential of
the reservoir (represented by a wide band model)
 for two values of the emf $\Phi=0.1,0.5$
(solid and dashed lines) and
two strengths of ISP $\sigma=0.05,0.85$ (thick black and thin red lines),
respectively for a ring with $N=200$ sites.}
\label{fig3}
\end{figure}
The behavior of $J_{dc}$ as a function of the chemical potential
is shown in Fig.\ref{fig3}. It is remarkable that a semicircular 
shape can be identified in the envelope of these functions. As mentioned
before, the density of states of a semi-infinite tight-binding chain
of bandwidth $W$, is proportional to a semicircle of radius $W/2$.
This motivates the comparison of the conductance of our device 
with
the conductance of     
a quantum dot which has two semi-infinite tight-binding
chains with hopping elements $T_h$ and
chemical potential $\mu_{\alpha}$ attached to its left and to its right,
through a hybridization amplitude $w_{eff}$. The latter corresponds to
the usual configuration employed in a Landauer-like calculation of
the conductance through a dot connected to a left and a right lead.
 Assuming that a small bias $V$ is
applied between the left and the right, the conductance of
such a system is given by 
a Landauer-like formula \cite{miwi}
\begin{equation}
G_{eff}=\frac{\partial J_{dc}}{\partial V}=
e \rho_0(\mu_\alpha) \Gamma,
\label{cond}
\end{equation}
where $\rho_0(\mu_\alpha)=\sqrt{4T_h^2-\mu_{\alpha}^2}/T_h^2$ is the
density of states of the semi-infinite chain and 
$\Gamma=w_{eff}^2 \rho_{eff}(\mu_{\alpha})/4 \pi$, being 
$\rho_{eff}(\mu_{\alpha})$ the density of states of the central system
dressed by the contact with the two chains. 

\begin{figure}
\includegraphics[width=65mm,angle=-90]{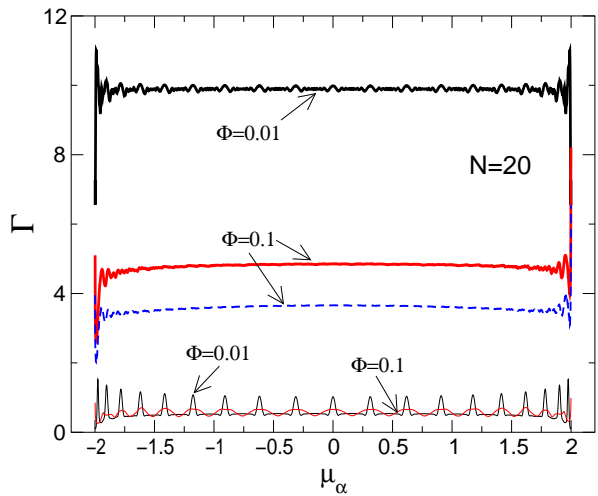}
  \caption{(Color online)
$\Gamma$ (proportional to the effective density of states
$\rho_{eff}(\mu_{\alpha})$ obtained by assuming a model for the 
conductance based on a dot coupled to two semi-infinite chains
for the case of a ring with $N=20$ sites and a wide band model for
the reservoir.
Thick and thin solid lines correspond to $\sigma=0.05$ and $\sigma=0.85$, 
respectively.
Black and red lines correspond to $\Phi=0.01$ and $\Phi=0.1$, respectively.
 The plot in blue dashed lines corresponds to $\Phi=0.1$ and a model
of reservoir with a semicircular density of states with $W=4$ and
$T_{\alpha}^2=0.1$. }
\label{fig4}
\end{figure}

In order to define the conductance for the device of Fig. \ref{fig0}, let us note
that the induced electric field is $E=\Phi/(Lc)$. Therefore, if we make the standart
assumption  that for
small $\Phi$ the current can be written as $J_{dc}=\sigma_c E$,
being $\sigma_c$ the conductivity, the conductance of the chain is defined from
$J_{dc}=G EL$.
Therefore, in units where ($c=1$), the conductance for small enough bias is  
\begin{equation}
G=\frac{J_{dc}}{\Phi}.
\label{cong}
\end{equation}

If we adopt the expression (\ref{cond}) as a {\em phenomenological}
model for the conductance of our system,  we 
obtain for $\Gamma$ the result shown in Fig. \ref{fig4}.
In the limit of weak coupling, this behavior is consistent
with an effective density of 
states approximately constant along the band-width.
Taking into account its sum rule, it should be
$\rho_{eff}(\mu_{\alpha})\sim \Theta(|\mu_{\alpha}|-4T_h)\pi/4T_{h}$. 
The large average value of $\Gamma$ indicates 
a large effective hopping from the dot to the semi-infinite chains
($w_{eff} \sim 12, 8$) for $\Phi=0.01, 0.1$, respectively. The
details of the model used for the reservoir do not influence the
final behavior of $\Gamma$. In fact, all the features in the
weak coupling regime observed within the wide band model for the
reservoir are also obtained with a reservoir with a semicircular density 
of states (cf plot in dashed lines of Fig. \ref{fig4}).  In the strong coupling
limit, $w_{eff}\sim 3$ is approximately the same for both fields and a fine
structure of $N$ wide peaks is distinguished in $\rho_{eff}(\mu_{\alpha})$.
The effective parameters $\Gamma$ and $w_{eff}$
do not have any straightforward significance in the context of our
original model and the parallel between the driven ring and 
the dot connected to semi-infinite chains is purely heuristic.
However, it is remarkable 
the similar behavior of the conductance of
the two devices. 
 It suggests that although the chain that forms
the ring is made up by a discrete chain with a finite number of sites, 
the dressing due to the mixing of a large number
of frequencies tend to produce a structure such that  any site
along the ring would feel as if placed between
two semi-infinite leads.
 In spite of 
this resemblance, the quantitative properties of the conductance are different
from the one which would be measured in a Landauer device with the
chain forming the ring
stretched and placed between two reservoirs, as already discussed in 
\cite{lili2}. In fact, neither the effective
 density of states at the site between the two
effective leads nor the hopping element $w_{eff}$ 
correspond to the configuration of a non-interacting
site coupled to two leads through $T_h$ as in the original
chain.

\subsection{The ring with a two-band structure}
As mentioned in the introduction, one of our motivations is to analyze
the behavior of the dc-current in a disordered system in order to 
explore the possibility of `resistive' behavior in the limit
of vanishing dissipation. 
We found it instructive to study
first an intermediate situation where the potential profile
is $\epsilon_{2n}=\epsilon_e, \; \epsilon_{2n-1}=\epsilon_o, n=1,\ldots,N/2$.
This structure defines two bands separated by an energy gap
which are the basic ingredient to discuss the effect of Zener 
tunneling.

\begin{figure}
\includegraphics[width=65mm,angle=-90]{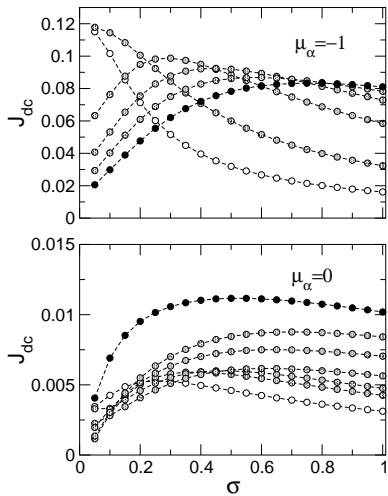}
  \caption{dc-current as a function of $\sigma$ for
two different chemical potentials in
a ring of $N=200$ sites with a two-band structure defined by 
$\epsilon_{o}=0, \; \epsilon_{e}=0.2$ and a reservoir with a constant
density of states. Different plots
correspond to $\Phi=0.05,0.1,0.2,0.3,0.4,0.5$ (upper panel)
and  $\Phi=0.2,0.3,0.4,0.5,0.6,0.7,0.8$ (lower panel). }
\label{fig5}
\end{figure}

\begin{figure}
 \epsfxsize=3.5in
  \epsffile{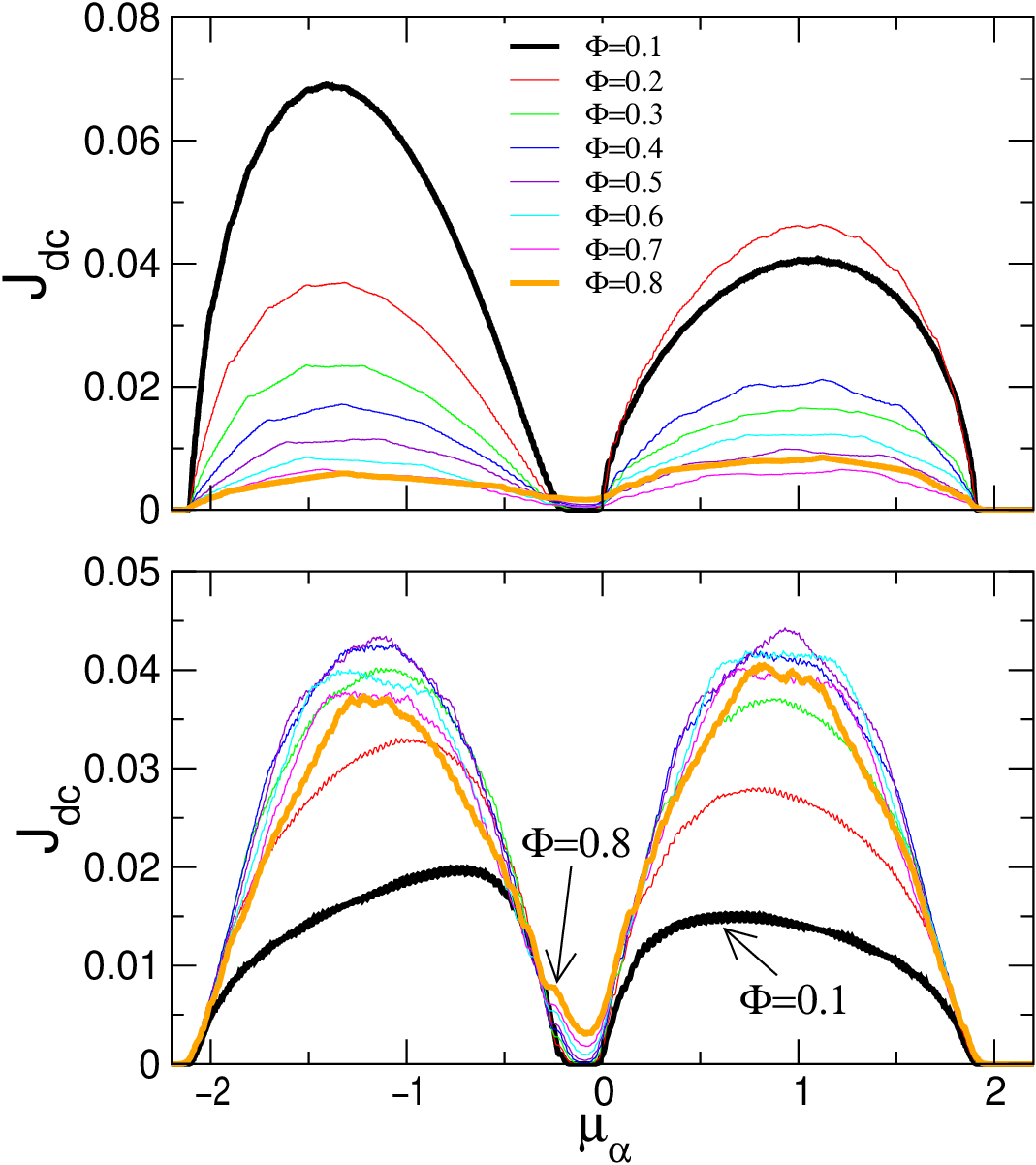}
  \caption{(Color online) dc-current as a function the chemical potential for 
a ring  with $N=200$ sites and a two-band structure defined by 
$\epsilon_{o}=0, \; \epsilon_{e}=0.2$ and a reservoir with a constant
density of states. The upper and lower panels correspond to  weak 
($\sigma=0.05$) and strong ($\sigma=0.85$) coupling regimes.
Different plots correspond to different fields
 $\Phi=0.1,0.2,0.3,0.4,0.5,0.6,0.7,0.8$. The ones corresponding to the
lowest and the highest fields are plotted in thick black and orange lines,
respectively.}
\label{fig6}
\end{figure}

The dc-component of the current as a function of $\sigma$
is shown in Fig. \ref{fig5} for a profile with $\epsilon_e=0.2,\;\epsilon_o=0$ 
and chemical potentials of the reservoir within the
lower band (upper panel) and within the gap (lower panel). 
In the first case, a behavior similar to that found in the
clean one-band model (cf Figs \ref{fig1},\ref{fig2}) is observed. Namely,
a maximum of the current that shifts to lower $\sigma$ as $\Phi$ 
decreases. The shift depends, however, slower with $\Phi$
in the present case:
taking as a reference the plot corresponding to $\Phi=0.2$, we see
that the maximum in the two-band case takes place at $\sigma \sim 0.2$
while in the clean ring $J_{dc}$ remained growing for the lowest
$\sigma$ used in our calculations.
The comparison of the plots of the
 upper panel of Fig. \ref{fig6} with those of the right
panel of  Fig. \ref{fig2} shows that the
magnitude of the maxima in the weak dissipation limit is
smaller in the two-band ring than in the case of the 
clean ring with the same number of sites in contact to a reservoir 
with the same characteristics and with the same chemical potential.
The behavior of $J_{dc}$ as a function of the chemical potential
shown in Fig. \ref{fig6}
shows that this is also the case for other values of the chemical
potential and dissipation strengths. The smaller current for 
chemical potentials within the lower conduction band, indicates that
the presence of an energy gap in the structure of energy levels
of the chain contributes to a less efficient mixing
of weights of the Green function at different frequencies relative
to the corresponding one in the clean one-band case. 

The effect of ISP on the magnitude of the current through the
gap is shown in the lower panel of Fig. \ref{fig5}. Within the weak coupling 
regime, the current is vanishingly small even for fields significantly
larger than the energy gap ($ \sim |\epsilon_o-\epsilon_e|$). This indicates 
that Zener tunneling alone (ie without dissipation) is not enough to 
generate a dc-response. 
    
The behavior of $J_{dc}$ as a function of the chemical potential is
shown in Fig \ref{fig6}. An important asymmetry is observed between the
upper and lower bands, which is more pronounced for low fields. 
This is an indication of the relevance of the
interband processes generated by the coupling of the electrons with the field. 
This discourages us from following the steps of the previous
section in trying to make a parallel with a device
based in two semi-infinite leads with the same band structure of the ring, 
since the latter would have two symmetric bands.  

\subsection{The ring with a random potential}
This final subsection is devoted to analyze a disordered
ring, with a random potential $\epsilon_l= \epsilon_w \gamma_l$,
being $-1 \leq \gamma_l \leq 1$, a random number.  
\begin{figure}
 \includegraphics[width=65mm,angle=-90]{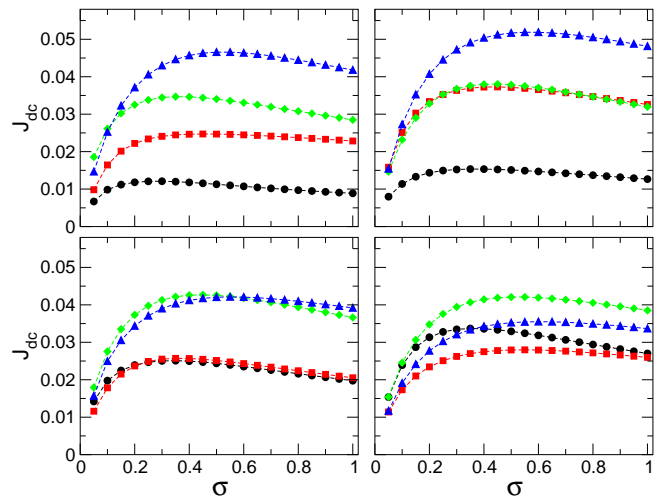}
  \caption{Color online) dc-current as a function of the strength of ISP for
a ring of $N=200$ sites with a random potential with amplitude
$\epsilon_w=0.2$ and a reservoir with a constant
density of states. Different panels correspond to different realizations
of the random potential. Black circles, red squares, green diamonds and
blue triangles correspond to $\Phi=0.2,0.3,0.4,0.5$, respectively.}
\label{fig7}
\end{figure}
The dc-current as a function of ISP strength
for a fixed chemical potential is shown in  Fig. \ref{fig7} for four different
realizations of the random potential. In all the cases a similar behavior
is found: the current exhibits a wide and mild maximum at intermediate
coupling and tend to zero in the limit of vanishing
ISP. The magnitude of $J_{dc}$ at the maxima is smaller
than in the two-band ring and significantly smaller than in the clean
one-band system. As mentioned in the previous subsection the
small current in the limit of weak coupling to the reservoir can be
interpreted by noting 
that the formation of energy gaps  in the structure of the
energy levels of the chain tend to break the resonant behavior
presented in the clean chain for small fields in the weak coupling
regime.   For small enough $\sigma$, the trend is $J_{dc} \propto \sigma$,
as in the clean limit.

\begin{figure}
\includegraphics[width=64mm]{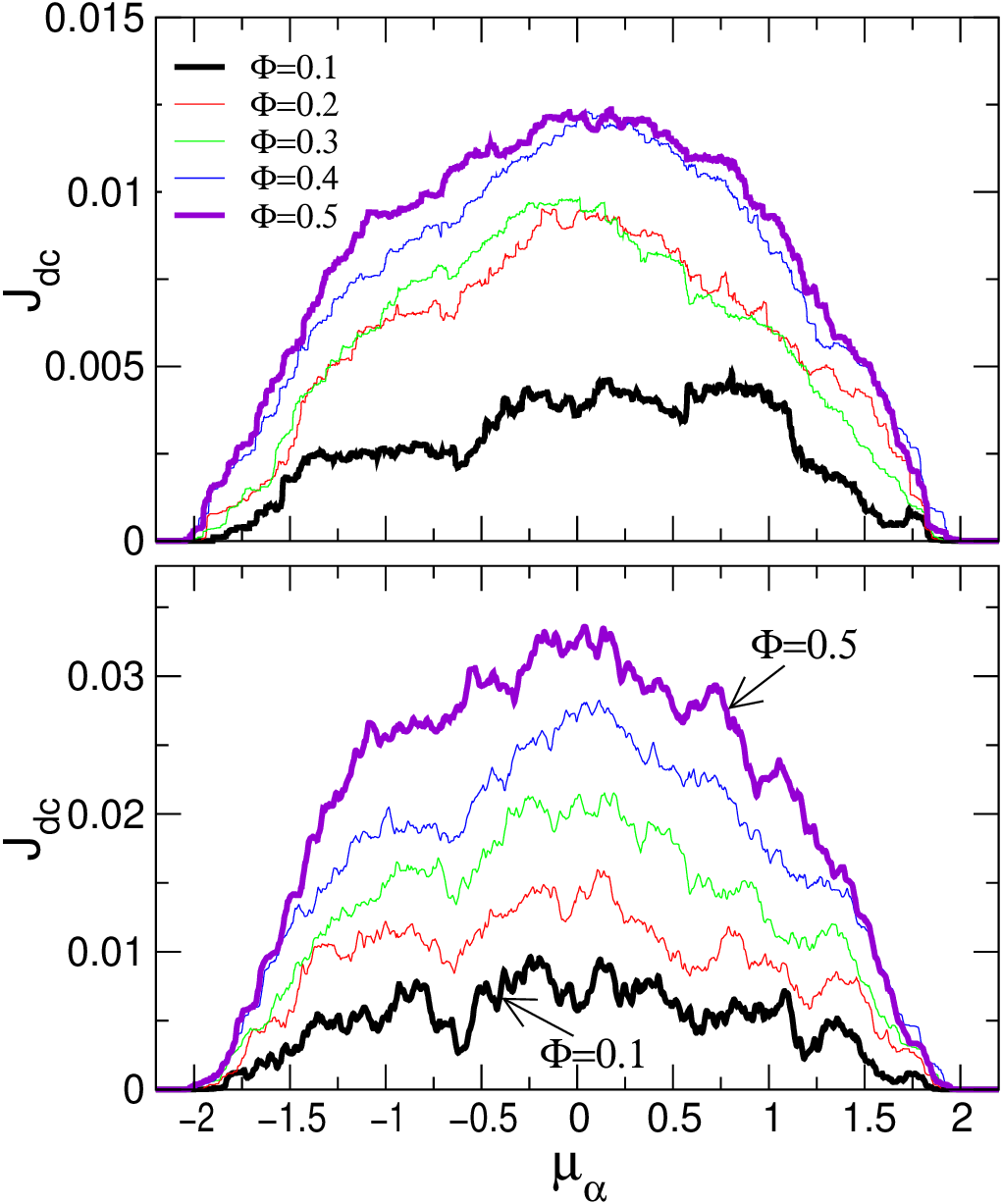}
  \caption{(Color online) dc-current as a function of the chemical potential for
a ring of $N=200$ sites with a random potential with amplitude
$\epsilon_w=0.2$ and a reservoir with a constant
density of states. 
The upper and lower panels correspond to  weak 
($\sigma=0.05$) and strong ($\sigma=0.85$) coupling regimes.
Different plots correspond to different fields
 $\Phi=0.1,0.2,0.3,0.4,0.5$. The ones corresponding to the
lowest and the highest fields are plotted in thick black and violet lines, respectively.}
\label{fig8}
\end{figure}

As a function of the chemical potential, the dc-current displays
fluctuations that become stronger in the strong coupling limit (see Fig. \ref{fig8}).
The fluctuations bear a close resemblance with those observed
in the behavior of 
the conductance of disordered quantum wires 
\cite{cf}. A systematic analysis of the 
probability distribution of 
conductance $G$ defined in (\ref{cong}) over several disorder realizations
as well as the behavior of $G$
as a function of the system length is left for the future.   

\section{Summary and conclusions}
We have studied the transport properties of a dissipative
 tight-binding
ring driven by means of a magnetic flux with a linear dependence 
in time. Dissipation is introduced by coupling the mesoscopic
ring to an external macroscopic system that plays the role of 
a reservoir of particles and energy.
We have extensively analyzed the behavior of the dc-component of
the current along the ring as a function of the strength of 
dissipation and of the chemical potential of the reservoir
for clean chains, chains with a two-band structure and an energy gap
and disordered chains.

We have analyzed the influence of the specific model assumed 
for the
reservoir and we conclude that this does not play any relevant role.
In the weak coupling regime, the structure of the energy levels
of the Hamiltonian describing the ring,
seems to play a relevant role. Such structure determines the
way in which contributions from different frequencies of the Green
function couple through the pumping term in Eqs. (\ref{a1}), (\ref{a2}).

In all the cases we expect a vanishing dc-current in the Hamiltonian
limit (vanishing ISP).   
For the case of the clean ring, we find that the current exhibits
a maximum as a function of the ISP strength that seems to scale 
with $\Phi$. For small enough
 $\Phi$ this maximum can lye very close to the limit of vanishing 
ISP. The range of fields where a linear behavior of
the dc-current as a function of $\Phi$ is observed depends on the
strength of ISP. It is wide for strongly coupled systems
and very narrow for weakly coupled ones. Remarkably, the behavior
of the conductance of the system as a function of the chemical potential
can be reproduced with a device consisting in
 a quantum dot with a constant density
of states and two semi-infinite tight-binding chains with the same
hopping parameter as the original one, attached to its left and right sides
through a very large hopping element.  

In the ring with a two-band structure, the behavior of the dc-current
within each of the two conducting bands is similar to that
observed in the one-band case. However, its magnitude 
is smaller and at fixed chemical potential and emf, the position
of its maximum as a function of the ISP strength
has a softer dependence with the emf. 
An important asymmetry is observed between the behavior of the 
current within the
upper and lower bands, while inelastic scattering is essential
to obtain a sizable magnitude of the current through the
gap, irrespectively the intensity of the induced emf.

The disordered ring can be seen as the N-band extension of the
two-band case. We have examined some typical realizations of the
random potential and observed that as a function of inelastic scattering,
the maximum in the dc-current takes place within the intermediate 
regime, becoming vanishingly small in the limit of weak coupling.
As a function of the chemical potential, the current displays 
fluctuations following patterns that depend
on the degree of inelastic scattering.

Our results are in agreement with the criteria based on symmetry 
arguments suggested for ratchet problems \cite{ser1,ser2,ser3}, 
according to which
a dc-component of the current is possible provided that the equations
of motion of the system are not invariant under symmetry operations
that change the sign of the time-dependent current. In the case of
the present problem the only possible symmetry that may
 introduce such an inversion, is time-reversal. This symmetry is
an exact one in the pure Hamiltonian limit where the ring is isolated
from the reservoir but it is immediately broken when the coupling to the 
reservoir is considered. 
Furthermore, we have found that the current grows linearly with the
parameter that characterizes the strength of the inelastic scattering. This behavior
has also been found in other pumped systems with broken time-reversal symmetry
\cite{leh}.
The behavior of the dc-current in the case
of the two-band and N-band (disordered) ring is also in agreement with  
the conclusions of Refs. \cite{lan1,lan2,gef1,gef2,brou} which 
have been devoted to argue  against the possibility of resistive
behavior caused by Zener tunneling alone. In our case we were able 
to solve the problem exactly and to examine details of
the behavior of the current for different strengths of
inelastic scattering. For this reason, we think that our results 
provide a robust support to the idea that dissipation is an essential
ingredient to obtain resistive behavior in this system. 

\section{Acknowledgments}
This work was in good part motivated by stimulating conversations
with Sergej Flach, whom the author thanks very specially, as well
as  Yuval Gefen and Victor Gopar for
useful correspondence and discussions.
The author is grateful to 
 Prof. Fulde for his kind hospitality at the MPIPKS-Dresden
and to Drazen Zanchi and Leticia Cugliandolo for their
hospitality at LPTHE, Paris, where this manuscript was finished.
LA also thanks the  support from
the Alexander von Humboldt Stiftung, CONICET, Argentina and
PICT 03-11609.

\end{document}